\documentstyle[emulateapj]{article}
\input{psfig}
\newcommand{\ls}{\raisebox{-0.5ex}{$\,\stackrel{<}{\scriptstyle 
\sim}\,$}}

\begin{document}

\title{Consequences of feedback from early supernovae for disk assembly}

\author{Priyamvada Natarajan}
\affil{Canadian Institute for Theoretical Astrophysics, 60 St. George Street, Toronto M5S A1, Canada}

\begin{abstract}
In this letter we examine the role of the first supernovae in
proto-galaxies, their role in feedback and the consequences for disk
assembly. Extending the picture proposed by Dekel \& Silk (1986), we
argue that energetic supernovae winds can expel baryons from all
proto-galaxies with varying degrees of efficiency. The fraction of
baryons retained and hence available to assemble into the baryonic
disk is therefore, a function of the central velocity dispersion of
the halo. Such a coupling of the baryonic component to the dark halo
leads to the following interesting consequence, a prediction for a
weak scaling of the zero-point of the Tully-Fisher relation or
alternatively, the mass-to-light ratio with the central velocity
dispersion of the halo. On applying to the case of the Milky Way halo,
this feedback mechanism implies: (i) that the Milky Way halo lost
$\sim\,10\%$ of its original gas content, and the gas mass lost is
roughly what is estimated for the mass in our X-ray halo consistent
with the X-ray background in the soft band; (ii) a range in the
inferred redshift of formation $z_f$, and the local baryon fraction
$f_b$ for the Milky Way that depends on the initial spin parameter
$\lambda_h$ of the halo. In a range of viable cold-dark matter
cosmological models, we find that for a low spin halo ($\lambda_h \sim
0.02$) - $z_f\,<\,1$, $f_b \sim 2\%$; for a median spin halo
($\lambda_h \sim 0.05$) - $z_f \sim 1 - 2.5$, $f_b \sim 5\%$; and for
a high spin halo ($\lambda_h \sim 0.2$) - $z_f \sim 4 - 8$, $f_b \sim
20\%$. The observationally determined ages for the oldest disk stars
in the Milky Way seem to rule out a low value for the spin
parameter. Given the shape of the spin distribution of halos obtained
in N-body simulations, while a high value of the spin parameter is not
very probable, it is interesting to note that if this is 
indeed the case for the Milky Way halo, then feedback processes can 
cause the local baryon fraction to differ significantly from the 
universal value.
\end{abstract}


\section{Introduction}

Despite tremendous recent progress both in the observations of
galaxies at high and low redshift and in the theoretical understanding
of their formation and evolution - there are still gaps in our
understanding of several of the physical processes that operate, most
notorious of them being `feedback'. The physics that couples baryons
to the dominant dark matter component is comparatively poorly
understood. There is compelling evidence from the many observed
correlations in galaxy properties for the existence of feedback
processes. In this work, we examine the feedback provided by the first
supernovae in proto-galaxies and the consequences for the
subsequent disk assembly in the halo. In $\S$ 2 the feedback process
is modelled, it is incorporated into a simple disk formation picture
in $\S$3, the consequences of the feedback are explored in $\S$4 with
emphasis on implications for the Milky Way disk and the conclusions
are presented in $\S$ 5.

\section{Feedback induced by early supernovae}

The very first SN to explode in a proto-galaxy have important effects
both in terms of setting into place feedback processes for subsequent
episodes of star-formation as well as causing the initial expulsion of
gas from the potential well. In this calculation, it is assumed that
the first episode of star formation in a proto-galaxy occurs well
prior to the assembly of the disk. This episode is expected to occur
in small, isolated and dense clumps in the cold phase of the gas.

The role of SN feedback and its implications for the evolution of a
galactic nucleus have been discussed by Dekel \& Silk (1986) [DS86
hereafter], Haehnelt \& Rees (1993) and Haehnelt (1994). Haehnelt \&
Rees (1993) estimate the relative importance of several issues in the
energetics: the impact on the heating-cooling balance due to the
thermal energy injected by the SN; the inhibition of the collapse
process produced by the radiation pressure from the ionizing radiation
as well as the mechanical energy of the expanding remnants; the effect
on further star formation in the nucleus due to the consequent complex
thermal structure of the gas, and the production and mixing of metals.
Semi-analytic models of galaxy formation (White \& Frenk 1991; Cole et
al. 1994; Kauffmann, White \& Guiderdoni 1993; Kauffmann 1996; Baugh
et al. 1997; Somerville \& Primack 1998) have also successfully
incorporated both the energy injection and mass loss due to
supernovae.

We focus here primarily on the dynamical effects of the first few
SN. While the radiation induced ionization effects act very locally,
and on small scales $\sim$ 100 pc, the energetic SN driven winds act
on much larger scales since typical velocities range from $\sim$ 5000
-- 10,000 km s$^{-1}$.  The mechanical effects of these winds are
invoked as a means to expel baryons from the potential well prior to
disk assembly. We show that this mechanism can modulate the fraction
of baryons that are retained in a proto-galactic potential well and
thereby naturally provides a variation in the fraction of available
baryons to form a disk that is a function of the velocity dispersion
of the halo.  Additionally, SN explosions during these early epochs
may expel preferentially high angular momentum material and,
therefore, the ejection could also aid in the subsequent efficient
disk assembly and the formation of central objects.

First, we briefly outline the calculation in DS86 on the expulsion of
gas from dwarf galaxies by supernovae driven winds. They computed the
full energetics and the evolution of the remnants through the two main
phases, the adiabatic expansion phase wherein the radiative cooling
losses are estimated assuming radiative $He^+$, Lyman-$\alpha$ and
$O_2$ line-cooling, with a cooling function
$\Lambda(T)\propto{{T^{-1}}\,\rho^2}$, for a metal-poor gas [$Z \sim
10^{-2}\,Z_{M_\odot}$]; and the expansion into the momentum conserving
snow-plow phase. Balancing the energy budget, they estimate a critical
velocity dispersion below which very little gas is retained in the
potential wells of dwarf-scale galaxy halos. This critical velocity
was found to be $v_{\rm crit}\,\ls\,100\,{\rm km\,s^{-1}}$. Here, we
take a more phenomenological approach to estimate the fraction of gas
that can be expelled from halos with velocity dispersions exceeding
this critical value. We assume that SN winds accelerate a variable
fraction of the gas to the halo's escape velocity. Unless the initial
mass function for this episode of star formation is considerably
different, it is expected that SN would produce a patchy ISM leading
to a complex relation between the fraction of baryons retained and the
depth of the potential well. The crucial point is that even for
massive halos a small fraction of the gas mass can be
lost/expelled. We assume in accordance with DS86 that $\sim$ 2\% of
the total SN energy output is available to drive the wind. The simple
energetics argument used here is as follows: an upper bound on the
fraction of gas mass lost can be computed by equating the escape
velocity of the halo with the bulk velocity of the expelled gas. The
total energy available per SN to drive the wind is given by:
\begin{eqnarray}
E_{\rm avail}\,=\,{2\,\times\,10^{49}}\,(\frac{\epsilon_{\rm
SN}}{0.02})\,(\frac{E_{\rm SN}}{10^{51}\,{\rm erg}})\,\,{\rm erg}.
\end{eqnarray}
The total number of SN is $N_{\rm SN}$, and energy balance implies that,
\begin{eqnarray} 
N_{\rm SN}\,{\epsilon_{\rm SN}}\,{E_{\rm SN}}\,=\,{\frac{M_l\,{v_{\rm
wind}}^2}{2}},
\end{eqnarray}
where $M_l$ is the gas mass that is lost and $v_{\rm wind}$ the wind
velocity. The escape velocity of the halo $v_{\rm esc}\, = 2 \sigma$;
therefore, for expulsion of the gas, we require that $v_{\rm
wind}^2\,\geq\,4\,\sigma^2$. The total number of SN is proportional to
all the mass that can potentially be turned into stars, i.e. the
baryonic mass of the halo $M_b\,\sim\,{f_b}\,M_h$, where $f_b$ is
constant and is of the order of 5\% in consonance with values
determined from the standard picture of light element production from
Big Bang Nucleosynthesis. Therefore, the total baryonic mass lost can be 
rewritten as,
\begin{eqnarray}  
M_l\,=\,\frac {2\,{N_{\rm SN}}\,{\epsilon_{\rm SN}}\,{E_{\rm SN}}}
{v_{\rm wind}^2}\,\propto \,{\frac {2\,{f_b}\,{M_h}\,{\epsilon_{\rm SN}}
\,{E_{\rm SN}}}{v_{\rm wind}^2}}
\end{eqnarray}
and on substituting the condition for gas expulsion, the fraction of 
baryonic mass lost $f_l$ is given by,
\begin{eqnarray} 
f_l\,\propto \,{\frac {{\epsilon_{\rm SN}}\,{E_{\rm SN}}}{2 \sigma^2}},
\end{eqnarray}
which can be translated into scaled units using the calibration 
obtained by DS86,
\begin{eqnarray} 
f_l\,=\,{5.0 \times 10^{-2}}\,({\frac {\epsilon_{\rm SN}}{0.02}})\,
({\frac {E_{\rm SN}}{10^{51}\,{\rm erg}}})\,
({\frac {\sigma}{200\,{\rm km\,s^{-1}}}})^{-2}.
\end{eqnarray}
By definition, it follows that the fraction of baryons retained
$f_r\,=\,(1 - f_l)$. Note that here we have assumed that the time
taken for the first stars to go SN, expand and expel the gas via their
winds is short compared to the time taken to assemble the disk. An 
attempt wherein feedback has been numerically implemented in a
tree-SPH simulation of disk galaxy formation is reported in the recent
work by Sommer-Larsen \& Vedel (1998). Nulsen \& Fabian (1997) have also
examined the role of the SN feedback in the context of the variation
of the Faber-Jackson relation for low-mass halos. 

\section{Incorporating feedback into simple disk formation models}

Mo, Mao \& White (1997) [MMW97] examined the formation of galactic
disks in the context of cosmological models with a view to explain and
understand the possible physical origin of their observed
properties. Here we follow their notation and incorporate the above SN
feedback into disk formation for 3 viable variants of the cold-dark
matter (CDM) family of cosmological models - standard CDM [SCDM]:
$\Omega_0 = 1$, $\Lambda = 0$; Open CDM [OCDM]: $\Omega_0 = 0.3$,
$\Lambda = 0$ and $\Lambda$CDM with $\Omega_0 = 0.3$, $\Lambda =
0.7$. The galactic-scale collapsed dark halo is modelled as an
isothermal sphere truncated at an outer radius $r_{200}$ where the
density is roughly 200 times the critical closure density $\rho_{\rm
crit}$.  A halo of a given mass $M_h$ is fully specified by its
velocity dispersion $\sigma$ and its redshift of collapse. The mass of
the halo is,
\begin{eqnarray}  
M_h\,=\,{\frac{2 {\sqrt 2} \sigma^3}{10\,G\,H(z)}}
\end{eqnarray}
where $H(z)$ is the Hubble constant at a given epoch $z$, defined as,
\begin{eqnarray}  
H(z)&=& \nonumber {100 h}\,[\Omega_{\Lambda} + (1 - \Omega_{\Lambda} - 
\Omega_{0})(1 + z)^2 \\ &+& \Omega_0 (1 + z)^3 ]^{1/2}\,\,{\rm km\,s^{-1}\,
{Mpc}^{-1}}
\end{eqnarray}
using the same notation as MMW97. 
The total baryonic mass available to form the disk, ${M_{\rm
disk}\,\propto\,f_r\,f_b\,M_{\rm halo}}$, therefore
from eqns. 6 and 7,
\begin{eqnarray}
M_{\rm disk}\,&=&\,{2.63\,\times\,10^{11}\,h^{-1}}\,[1 - {5.0 \times 10^{-2}}\,
({\frac{\sigma}{200\,{\rm km s^{-1}}}})^{-2}]\,
\nonumber \\ &\times&
({\frac
{f_b}{0.05}})\,({\frac {\sigma}{200\,{\rm
km\,s^{-1}}}})^3\,({\frac {H(z)}{H_0}})^{-1}\,\,M_{\odot}.
\end{eqnarray}
Initially the baryons and the dark matter have comparable torques but
dissipation causes differential settling of the baryons. We assume
that the total angular momentum of the disk $J_d$ is proportional to
the angular momentum of the halo $J_h$, $J_d\,=\,f_j\,J_h$. The
angular momentum of the disk is computed assuming a flat rotation
curve neglecting the inner-most region where the disk is
self-gravitating (models that take into account the back-reaction of
the halo have been worked out in detail by Blumenthal et al. (1986),
Dalcanton et al. (1997) and MMW97). The spin parameter $\lambda_h$ of
the halo is a measure of the acquired angular momentum during the
collapse process, and is defined as $\lambda_h\,=\,J\,{E}^{\frac 1
2}\,G^{-1}\,{M}^{-{\frac 5 2}}$ where $J$, $E$ and $M$ are the total
angular momentum, energy and mass of the halo. Assuming that the disk
settles down to an exponential radial surface density profile, the
following expression can be obtained for the disk-scale length $r_d$
in terms of the relevant parameters,
\begin{eqnarray}
r_d\,&=&\,{8.17\,h^{-1}}\,({\frac {\lambda_h}{0.05}})\,({\frac
{f_j}{f_b}})\,[1 - {5.0 \times 10^{-2}}\,
({\frac{\sigma}{200\,{\rm km s^{-1}}}})^{-2}]^{-1}\, \nonumber \\ 
&\times&
({\frac {H(z)}{H_0}})^{-1}\,\,{\rm kpc}.
\end{eqnarray}
The Tully-Fisher (T-F) relation can be written as,
\begin{eqnarray}
\log\,[L_{\rm disk}]\,=\,\log\,[L_0]\,+\,{\beta}\,\log\,[{\frac
{\sigma}{200\,{\rm km\,s^{-1}}}}],
\end{eqnarray}
where $\beta$ is the Tully-Fisher index and $L_0$ the zero-point of
the relation. A recent determination by Giovanelli et al. (1997) in
the I-band, yields $\beta = 3$ and $L_0 =
{3.5\,\times\,10^{10}\,h^{-2}\,L_{\odot}}$.  Although the value of the
index $\beta$ does depend on the wave-band ($\beta = 2.5$ in the
B-band), the scatter is small (roughly 0.35 mags.) is magnitude
dependent and increases steeply with decreasing velocity width.
Assuming an efficiency ratio $\Upsilon$ for the conversion of gas in
the disk into stars, the disk mass $M_d$ (eqn. 8) can be translated
directly into a luminosity. Comparing this with the T-F relation, we
require,
\begin{eqnarray}  
M_{\rm
disk}\,\Upsilon^{-1}\,&=&\nonumber\,{2.63\,\times\,10^{11}\,h^{-1}}\,
[1 - {5 \times 10^{-2}}\,({\frac {\sigma}{200\,{\rm km s^{-1}}}})^{-2}]
\, \nonumber \\ &\times& ({\frac {f_b}{0.05}})\,({\frac
{\sigma}{200\,{\rm km\,s^{-1}}}})^3\, ({\frac
{H(z)}{H_0}})^{-1}\,{\Upsilon^{-1}}\, \nonumber \\ &=&\,{L_0}\,
({\frac {\sigma}{200\,{\rm km\,s^{-1}}}})^3.
\end{eqnarray}
There are now 2 possible ways to interpret the above equation, in
terms of (i) yielding a zero-point $L_0$ for the T-F relation that
{\bf does} depend on the depth of the potential or (ii)
if we require $L_0$ to be independent of $\sigma$, then it follows
that $\Upsilon$ is a weak function of the velocity dispersion of the
halo. Even if the T-F zero-point depends on $\sigma$, we find that it
is independent of the spin of the halo ${\bf \lambda_h}$ as first
pointed out by White (1987).

\section{Consequences of the feedback}

A stronger coupling between the velocity dispersion and disk mass
arises as a result of the feedback process. We make no claims with
regard to understanding why the T-F exponent is 3, it is assumed to be
so motivated by the observations. Any successful theory of galaxy
formation needs to account for both the slope of the relation as well
as the low scatter.

\subsection{Application to the Milky Way disk}

In the context of this feedback process, we can now examine what the
observed parameters of the disk of our Galaxy imply for its formation
history. The age of the disk stars in the Milky Way can be used as a
diagnostic to pin down the redshift of assembly. Also, constraints can
be obtained on the frequency of accretion events in our halo if the
bulk of the stars in the Milky Way disk assembled at $z\,>\,1$. The
thick disk stars have a relaxed exponential profile and were hence
probably in place quite early on. Moreover, since the relaxation
time-scale for stellar systems is long compared to their typical
dynamical times, these stars most probably formed in situ. Chemical
evolution models of our galaxy seem to indicate that a constant
star-formation rate is not a bad description (Wyse \& Gilmore (1995)),
implying that half the stellar mass of our Galaxy probably assembled
well before $z = 1$.

An early generation of stars (progenitors of these SN) also provides a
natural solution to the so-called G-dwarf problem, i.e. the paucity of
metal poor stars in our Galaxy that closed-box chemical evolution
models predict in abundance (Pagel 1989). Observationally determined
parameters for our Galaxy are: a disk mass
$M_{MW}\,\sim\,6\,\times\,10^{10}\,M_{\odot}$, velocity dispersion
$\sigma\,\sim\,155$ km s$^{-1}$ and a disk-scale length
$r_d\,\sim\,3.5$ kpc. Firstly, in accordance with this prescription,
the predicted gas fraction lost by the Milky Way halo,
$f_l\,\sim\,10\%$. Substituting the appropriate values in eqns. 8 and
9, assuming maximal efficiency for disk formation $f_b = f_j$ and
$H_0\,=\,50$ km s$^{-1}$ Mpc$^{-1}$, we obtain the following relations:
\begin{eqnarray}
({\frac {\lambda_h}{0.05}})\,({\frac {H(z)}{H_0}})^{
-1}\,=\,{0.254};\,\,\,\,({\frac {f_b}{0.05}})\,
({\frac {H(z)}{H_0}})^{-1}\,=\,{0.267}.
\end{eqnarray}
Combining these two bounds yields,
\begin{eqnarray}
({\frac {\lambda_h}{0.05}})\,({\frac
{f_b}{0.05}})^{-1}\,=\,{1.054}.
\end{eqnarray}
The constraint above can be interpreted as providing bounds on the
value of the local baryon fraction depending on the value of the
initial spin parameter of the Milky Way halo. Numerical simulations
and analytic estimates seem to indicate that the spin distribution of
halos $P(\lambda)$ can be well-fit by a lognormal with a mean value of
0.05 (Ryden 1988; Warren et al. 1992). If the spin of the Milky Way
halo was low $\lambda_h \sim 0.02$ then eqn. (13) implies that the
local baryon fraction ($f_b \sim 2\%$) is lower by a factor of two 
compared to the universal value; a mean value for the spin parameter
yields $f_b \sim 5\%$ in agreement with the universal value whereas
a high value of the spin parameter ($\lambda_h \sim 0.2$) implies
$f_b \sim 20\%$ which is much higher than the universal value but is
in good agreement with the value inferred from X-ray studies of the
Intra-cluster medium in clusters of galaxies. 

A feedback scenario such as the one suggested in this work - in which
only 10\% of the gas is ejected has interesting implications for the
origin of the extended X-ray halo around our Galaxy. The limit on the
gas mass expelled from the halo of our Galaxy predicted by this 
model is in good agreement with the mass
of the hot corona of our Galaxy (Bregman 1980; Kerp et al. 1998)
consistent with the soft X-ray Background.  Given values of $f_b$ and
$\lambda_h$ eqn. (12) also then provides naturally an
estimate for the formation redshift for our Galaxy - (i) for a low
spin, $f_b \sim 2\%$ $z_f\,<\,1$, (ii) a mean spin, $f_b \sim 2\%$
$z_f\,\sim\,[1.0 - 2.5]$, and (iii) for a large spin, $f_b \sim
20\%$ $z_f\,\sim\,[4.0 - 8.0]$, for the 3 cosmological models 
considered here [as shown in Fig. 1]. The ages of the oldest disk 
stars in our Galaxy (Wyse 1997), seem to be consistent with a 
high-redshift of formation, thereby ruling out a low spin parameter
for the Milky Way halo.

\section{Discussion and Conclusions}

In this letter we have examined the consequences of feedback from an
early episode of SN explosions in proto-galaxies. An extension of the
mechanism proposed by Dekel \& Silk (1986), to higher velocity
dispersion halos yields a modulation of the baryon fraction available
for disk formation that depends on the depth of the potential well.
This feedback applied to simple prescriptions for disk formation
implies that the either the zero-point of the Tully-Fisher or the
mass-to- light ratio is weakly dependent on the velocity dispersion of
the halo. The implications of this scenario for the Milky Way disk are
that approximately 10\% of the original gas that fell in was expelled
prior to disk assembly, a high-redshift of formation which is in good
agreement with the ages of the oldest thick disk stars is inferred for
average to high values of the spin parameter of the Milky Way halo with
a local baryon fraction that ranges from the universal value of 5\% to
a value of 20\% which is what is inferred from X-ray studies of 
the Intra-cluster medium in clusters of galaxies.

\acknowledgements

Useful conversations with Martin Haehnelt, Martin Rees, Steinn
Sigurdsson, Joe Silk, Rachel Somerville and Rosie Wyse are 
gratefully acknowledged. PN thanks the anonymous referee for 
very constructive comments.

\newpage

\begin{figure}
\plotone{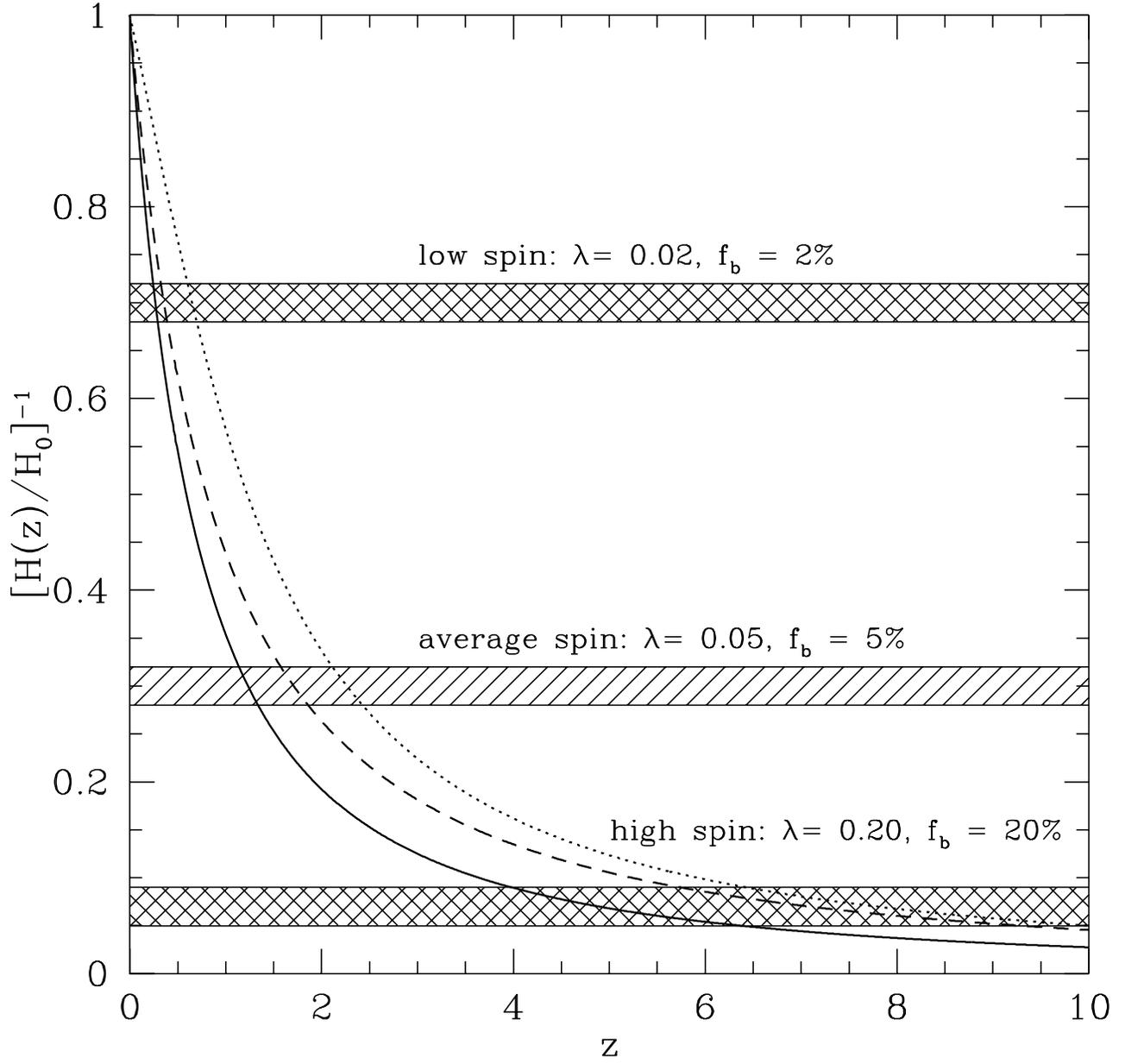}
\caption{The constraint on the redshift of assembly for the Milky Way disk
obtained from early SN feedback is marked by the hatched region. 
The solid curve is the SCDM model, the dashed curve OCDM and the 
dotted curve is for the $\Lambda$CDM model.}
\end{figure}

\end{document}